\documentstyle[12pt,epsf]{article}
\baselineskip=\normalbaselineskip
\ifx\TwoupWrites\UnDeFiNeD\else\target{\magstepminus1}{11.3in}{8.27in}
	\source{\magstep0}{7.5in}{11.69in}\fi
\newfont{\fourteencp}{cmcsc10 scaled\magstep2}
\newfont{\titlefont}{cmbx10 scaled\magstep3}
\newfont{\authorfont}{cmcsc10 scaled\magstep1}
\newfont{\fourteenmib}{cmmib10 scaled\magstep2}
	\skewchar\fourteenmib='177
\newfont{\elevenmib}{cmmib10 scaled\magstephalf}
	\skewchar\elevenmib='177
\makeatletter
\newcommand\nonsequentialeqnum{
	\@addtoreset{equation}{section}
	\def\theequation{\arabic{section}.\arabic{equation}}}
\newif\ifp@bblock  \p@bblocktrue
\newcommand\nopubblock{\p@bblockfalse}
\newcommand\topspace{\hrule height 0pt depth 0pt \vskip}
\newcommand\p@bblock{\begingroup \tabskip=\hsize minus \hsize
	\baselineskip=1.5\ht\strutbox \topspace-2\baselineskip
	\halign to\hsize{\strut ##\hfil\tabskip=0pt\crcr
	\the\Pubnum\crcr\the\date\crcr}\endgroup}
\renewcommand\titlepage{\ifx\TwoupWrites\UnDeFiNeD\null\vspace{-1.7cm}\fi
\vskip0.6cm
	\ifp@bblock\p@bblock \else\hrule height 0pt \relax \fi}
\makeatother
\newtoks\date
\newtoks\Pubnum
\newtoks\pubnum
\Pubnum={
FIT-HE-96-81 \crcr
KYUSAN-TH-381 \crcr
August 1996\crcr
hep-th/xxxxxxxx
}
\newcommand{\frontpageskip}{\vspace{12pt plus .5fil minus 2pt}}
\renewcommand{\title}[1]{\frontpageskip
	\begin{center}{\titlefont #1}\end{center}\par}
\renewcommand{\author}[1]{\frontpageskip\par\begin{center}
	{\authorfont #1}\end{center}
	\nobreak
	}

\newcommand{\address}[1]{\par\begin{center}{\sl #1}\end{center}\par}

\renewcommand{\thanks}[1]{\footnote{#1}}
\renewcommand{\abstract}{\par\frontpageskip\centerline{\fourteencp Abstract}
	\vspace{8pt plus 3pt minus 3pt}}
\begin{document}
\pubnum{93-37}
\titlepage

\renewcommand{\thefootnote}{\fnsymbol{footnote}}
\title{
Effective Potential in the String Induced Action \\
}

\author{
K.\ Ghoroku${}^{1\,}$\thanks{
e-mail address: gouroku@dontaku.fit.ac.jp},
K.\ Kaneko${}^{2\,}$\thanks{
e-mail address: kaneko@daisy.te.kyusan-u.ac.jp}
}

\address{
${}^1$
Fukuoka Institute of Technology\\
Wajiro, Higashi-ku, Fukuoka 811-02, Japan \\
${}^2$
Kyushu Sangyo University\\
Matsukadai, Fukuoka 813, Japan\\
}

\renewcommand{\thefootnote}{\arabic{footnote}}
\setcounter{footnote}{0}
\newcommand{\cleqn}{\setcounter{equation}{0} \indent}
\renewcommand{\theequation}{\thesection.\arabic{equation}}
\newcommand{\beqa}{\begin{eqnarray}}
\newcommand{\eeqa}{\end{eqnarray}}
\newcommand{\eq}[1]{(\ref{#1})}

\begin{abstract}

The dynamical properties of the gauge theory of Born-Infeld type 
action, which is expected as the high-energy effective theory,
are investigated by adding a complex scalar field to
this gauge system. Especially the Coleman-Weinberg mechanism is addressed
in this theory.

~\\
~\\

\end{abstract}

\newpage

\newcommand{\scs}{\setcounter{equation}{0} \setcounter{section}}
\def\req#1{(\ref{#1})}
\setcounter{footnote}{0}

\section*{1. Introduction}\scs{1}

A non-linear form of gauge field action, the Born-Infeld (BI) type,
has been obtained from string theory \cite{ftsey}, and it is recently
attracting many attentions as an effective action of D-brane 
\cite{schm},\cite{tse} in order to study the non-perturbative aspect
of string theory. Even if we do not relate the BI action 
directly to the D-branes, 
this action is interesting as a high-energy form of the gauge action.
The reason that BI action can be regarded as the high-energy form is that
it includes the low-energy form, $F_{\mu\nu}F^{\mu\nu}$, in the lowest order
of series expansions of BI action with respect to $\alpha'$, which is
the mass-scale of the order of (Planck Mass)$^{-2}$. 

Gauge fields had played an important role in the various theories, and they
are cleared through many analyses in terms of the low-energy action, 
$F_{\mu\nu}F^{\mu\nu}$. The results obtained up to now are of course
valid in the low-energy region. And we might expect that these results are
also being valid in the high-energy region.
However we must use the BI action in order to see whether
the dynamics at low-energy is still valid 
at very high energy, near the Planck mass-scale, or not.

Our purpose here is to examine the validity of the Coleman-Weinberg
mechanism in the high-energy region where the BI action should be used. 
The analysis
is parallel to \cite{cw}; Consider the gauge system with a complex scalar
field, and calculate the effective potential of the scalar field taking into
account of the radiative correction in order to see the spontaneous breakdown
of the gauge symmetry. In this calculation, we introduce several
auxiliary fields to keep the non-linearity of the gauge action. 
This technique would be useful to see the duality problem of the 
corresponding D-brane action.

\section*{2. Born-Infeld action}\scs{2}

There might be two equivalent ways to obtain the BI gauge action.
One way \cite{clny} is based on the BRST invariance of the 
super-string theory
where the boundary states are introduced in order to calculate
the loop corrections of string. This method is also useful to 
obtain the D-brane action \cite{schm},

The other way is the path-integral method given in \cite{ftsey}.
We briefly review this method. The general form of the effective action, 
which is obtained from the interacting system of open and closed strings,
is written as follows,

\begin{eqnarray}
  S_{\rm eff}(A_{\mu}) &=& \sum_{\chi=1,0,\cdots}{\rm e}^{\sigma\chi}
         \int dg_{ab}dx^{\mu}\rm{e}^{-I_2-I_1}, \label{eq:a6} \\
  I_2 &=& {1 \over 4\pi\alpha'}\int_{M}d^2z \sqrt{g}g^{ab}
           \partial_ax^{\mu}\partial_bx^{\mu}, \label{eq:a7} \\
  I_1 &=& \int_{\partial M}dt i\dot{x}^{\mu}A_{\mu}, \label{eq:a8}
\end{eqnarray}
where $\chi$ denotes the Euler number, and
the background of the closed string is taken as 
$G_{\mu\nu}=\delta_{\mu\nu}$ and other fields are zero.
As for the open string, only the gauge fields $A_{\mu}$
are retained. This action can be calculated for
the various number of $\chi$ or the genus, and the BI action 
is obtained for the disc amplitude ($\chi=1$) by assuming constant
$F_{\mu\nu}$.

The path integral is performed for $x^{\mu}$ by separating it to
the center of mass of the string $y^{\mu}$ and the fluctuation $\xi^{\mu}$,
$x^{\mu}=y^{\mu}+\xi^{\mu}$. The essential part of the integration is
the one of $\xi^{\mu}$ on the boundary $\partial M$ of the world surface.
In performing this integration, the
important point is the following expansion of the boundary
action,
\begin{equation}
 \int dt\dot{\xi}^{\mu}A_{\mu}(y+\xi)={1 \over 2}F_{\mu\nu}(y)\int dt
            \dot{\xi}^{\mu}\xi^{\nu}
            +{1 \over 3}\partial_{\lambda}F_{\mu\nu}(y)
            \int dt\dot{\xi}^{\mu}\xi^{\nu}\xi^{\lambda}+\cdots.
                            \label{eq:a1}
\end{equation}
Then we obtain the gaussian form for $\xi$ if we assume $F_{\mu\nu}=$
constant, and the integration can be performed exactly. The result is
obtained as follows,
\begin{equation}
 S_{\rm eff}(A_{\mu}) \ =\  g_0^{-2}\alpha'^{-d/2}
   \int d^{d}y\ \ {\sqrt{\det(G+2\pi\alpha'F)}} \ ,
 \label{eq:a2}
\end{equation}
where $G_{\mu\nu}=\delta_{\mu\nu}$, $g_0=\rm{e}^{-\sigma/2}$ and $d$
denotes the space-time dimension.

For the one-loop amplitude ($\chi=0$), urtra-violet
divergence proportional to the disc-amplitude appears, and
it could be absorbed into the coupling constant, $g_0$, in the tree 
amplitude. 

In this way, we obtain the BI action from string theory.
It should be noticed that this action can be expanded by the series of 
$\alpha'$, and we obtain the low-energy effective action,
$F_{\mu\nu}F^{\mu\nu}$, when the higher order terms of $\alpha'$ are
neglected. While these terms become important at high energy, and
they must be summed up. 
The higher order terms of $\alpha'$ also appear if we do not
restrict to the case of $F_{\mu\nu}=$ constant. 
In this case, the terms of
higher power of $\xi$ in (\ref{eq:a1}) remain, and we obtain
the $\alpha'$ series which are different from the one coming from
the expansion of BI action.
So the BI action can be considered
as a special sum of the $\alpha'$ series expansion, and this form
becomes exact for constant $F_{\mu\nu}$. 
However we notice that the BI action is
the lowest part with respect to the string
coupling constant $g_0$. Here we concentrate on the small $g_0$ region,
then the above BI action can be regarded as an effective action of a
gauge theory at high energy. 
In this paper, we use it in the analysis below
by assuming that $F_{\mu\nu}$ is slowly varying.

\section*{3. Effective potential of charged scalar field}\scs{3}

   In this section, we discuss Coleman-Weinberg mechanism in string induced action coupled with a massless charged scalar fields. The Euclidean action is given as
\begin{equation} 
S_{d}= \int d^{d}x {\cal L}(x), 
\end{equation}
where the Lagrangian density is given by
\begin{equation} 
{\cal L}(x)= \sqrt{{\rm det}(\delta_{\mu\nu}+2\pi\alpha' F_{\mu\nu})} + \frac{1}{2}\mid D_{\mu}\varphi\mid^{2} + \lambda (\mid\varphi\mid^{2})^{2}, 
\end{equation}
$$ F_{\mu\nu}=\partial_{\mu}A_{\nu}-\partial_{\nu}A_{\mu},$$
$$ D_{\mu}= \partial_{\mu} - ieA_{\mu}.$$
Here, we assume the $d$-dimensional space-time to have Euclidean signature. We have two small coupling constants $\lambda$ and $e$.
We now add the Lagrange multiplier $\Lambda^{\mu\nu}$ term 
\begin{equation} 
S_{d}= \int d^{d}x [ \sqrt{{\rm det}(\delta_{\mu\nu}+2\pi\alpha' \tilde{F}_{\mu\nu})} + \frac{1}{2}\mid D_{\mu}\varphi\mid^{2} + \lambda (\mid\varphi\mid^{2})^{2} + \frac{1}{2}i\Lambda^{\mu\nu}(\tilde{F}_{\mu\nu}-2\partial_{\mu}A_{\nu}) ], \end{equation}
where $\Lambda^{\mu\nu}$ is antisymmetric, 
$\Lambda^{\mu\nu}=-\Lambda^{\nu\mu}$.
The terms, which include the field $A_{m}$, are expressed as
$$ -i\Lambda^{\mu\nu}\partial_{\mu}A_{\nu}+\frac{1}{2}\mid (\partial_{\mu}-igA_{\mu})\varphi\mid^{2}
\bar{P}^{(1)}_{\mu\nu\rho\sigma}\
 + \frac{1}{2}\gamma^{2}(\bar{P}^{(2)} + \bar{P}^{(1)})_{\mu\nu\rho\sigma}
\qquad\qquad\qquad\qquad\qquad 
$$
$$=\frac{1}{2}e^{2}\mid\varphi\mid^{2}[A_{\mu}+\frac{i}{e^{2}\mid\varphi\mid^{2}}\{ (\partial_{n}\Lambda^{\nu\mu})+\frac{1}{2}e(\varphi^{\dagger}\partial_{\mu}\varphi-\partial_{\mu}\varphi^{\dagger}\varphi ) \} ]^{2}$$
\begin{equation} 
 +\frac{1}{2e^{2}\mid\varphi\mid^{2}}\{ (\partial_{\nu}\Lambda^{\nu\mu})+\frac{e}{2}(\varphi^{\dagger}\partial_{\mu}\varphi-\partial_{\mu}\varphi^{\dagger}\varphi )\}^{2}+\frac{1}{2}\mid\partial_{\mu}\varphi\mid^{2}.
\end{equation}
Substituting these terms into the action and integrating $A_{m}$, we have
$$ S_{d}= \int d^{d}x [ \sqrt{{\rm det}(\delta_{\mu\nu}
     +2\pi\alpha' \tilde{F}_{\mu\nu})} 
     + \frac{1}{2}i\Lambda^{\mu\nu}\tilde{F}_{\mu\nu} 
\qquad\qquad\qquad\qquad\qquad
$$
\begin{equation} 
 +\frac{1}{2e^{2}\mid\varphi\mid^{2}} \{ (\partial_{\nu}\Lambda^{\nu\mu})+\frac{e}{2}(\varphi^{\dagger}\partial_{\mu}\varphi-\partial_{\mu}\varphi^{\dagger}\varphi ) \}^{2} +\frac{1}{2}\mid\partial_{\mu}\varphi\mid^{2}+ \lambda (\mid\varphi\mid^{2})^{2} ].
\end{equation}
The BI part is rewritten with the help of an auxiliary scalar field $v$:
$$ {\rm exp}( -\int d^{d}x \sqrt{{\rm det}(\delta_{\mu\nu}
        +2\pi\alpha' \tilde{F}_{\mu\nu})} ) 
      \qquad\qquad\qquad\qquad
$$
\begin{equation} 
 = \int {\cal D}v {\rm exp} \{ -\int d^{d}x [ \frac{1}{2v^{2}}({\rm det}(\delta_{\mu\nu}+2\pi\alpha' \tilde{F}_{\mu\nu})) + \frac{1}{2}v^{2} ] \}.
\label{eqn:BIpart}
\end{equation}
Here, ${\rm det}(\delta_{\mu\nu}+2\pi\alpha' \tilde{F}_{\mu\nu})$ is expressed as
\begin{equation} 
 {\rm det}(\delta_{\mu\nu}+2\pi\alpha' \tilde{F}_{\mu\nu})= 1+\frac{1}{2}(2\pi\alpha')^{2}\tilde{F}_{\mu\nu}^{2}, \hspace{1cm} {\rm for}\hspace{0.3cm} d=2,3 
\label{eqno:d=2,3}
\end{equation}
\begin{equation} 
 =1+\frac{1}{2}(2\pi\alpha')^{2}\tilde{F}_{\mu\nu}^{2}+(\frac{1}{2}(2\pi\alpha')^{2}\varepsilon_{ijk}\tilde{F}_{0i}\tilde{F}^{jk})^{2}, \hspace{0.5cm} {\rm for}\hspace{0.3cm} d=4 
\label{eqn:d=4}
\end{equation}
where $i,j,k=1 \sim 3$.
\subsection{\bf $d=4$}

 In the case of $d=4$, third term of eq.(\ref{eqn:d=4}) makes the integration
about the fields $\tilde{F}_{mn}$ difficult. However, if we take $\alpha'$ 
as an expansion parameter, third term of eq.(\ref{eqn:d=4}) is higher order, 
$\alpha'^{4}$. We can thus neglect the third term of eq.(\ref{eqn:d=4}) for
small $\alpha'$.
\footnote{It is possible to consider that this term is removed by assuming
$\varepsilon_{ijk}\tilde{F}_{0i}\tilde{F}^{jk}=0$. If one define electric fields $E_{i}=\tilde{F}_{0i}$ and magnetic fields $H_{i}=\varepsilon_{ijk}\tilde{F}^{jk}/2$, this means $\vec{E}\cdot\vec{H}=0$. }
Then, the action is 
$$ S_{4}= \int d^{4}x [ (\tilde{F}_{\mu\nu}+\frac{iv}{2}\Lambda^{\mu\nu})^{2} + \frac{1}{4}v^{2}(\Lambda^{\mu\nu})^{2} + \frac{1}{2v^{2}}+\frac{1}{2}v^{2} 
    \qquad\qquad\qquad\qquad
$$
\begin{equation}
 +\frac{1}{2e^{2}\mid\varphi\mid^{2}} \{ (\partial_{\nu}\Lambda^{\nu\mu})+\frac{e}{2}(\varphi^{\dagger}\partial_{\mu}\varphi-\partial_{\mu}\varphi^{\dagger}\varphi ) \}^{2} +\frac{1}{2}\mid\partial_{\mu}\varphi\mid^{2}+ \lambda (\mid\varphi\mid^{2})^{2} ], 
\end{equation}
where we used the transformation $\tilde{F}_{\mu\nu}\rightarrow \frac{v}{\pi\alpha'}\tilde{F}_{\mu\nu}$.
Performing the gaussian functional integral about $\tilde{F}_{\mu\nu}$, we obtain the following Lagrangian density,
$$ \tilde{\cal L}(x)= \frac{1}{4}v^{2}(\Lambda^{\mu\nu})^{2} + \frac{1}{2v^{2}}+\frac{1}{2}v^{2} 
+\frac{1}{2e^{2}\mid\varphi\mid^{2}} \{ (\partial_{\nu}\Lambda^{\nu\mu})+\frac{e}{2}(\varphi^{\dagger}\partial_{\mu}\varphi-\partial_{\mu}\varphi^{\dagger}\varphi ) \}^{2} $$
\begin{equation}
+\frac{1}{2}\mid\partial_{\mu}\varphi\mid^{2}+ \lambda (\mid\varphi\mid^{2})^{2} . 
\label{eqn:Lagrangian}
\end{equation}

  We will here organize perturbation theory in the form of loop expansion, and derive the effective potential with respect to $\varphi_{c}$. 
Perturbing around the classical field $\varphi_{c}$, we define a shifted field
as follows:
\begin{equation}
 \delta\varphi= \varphi - \varphi_{c}, 
\end{equation}
where we assume that $\varphi_{c}$ has a real value.
Then, the terms with respect to $\varphi$ in eq.(\ref{eqn:Lagrangian}) are expanded as
$$\frac{1}{2e^{2}\mid\varphi\mid^{2}} \{ (\partial_{\nu}\Lambda^{\nu\mu})+\frac{e}{2}(\varphi^{\dagger}\partial_{\mu}\varphi-\partial_{\mu}\varphi^{\dagger}\varphi ) \}^{2}+\frac{1}{2}\mid\partial_{\mu}\varphi\mid^{2}+ \lambda (\mid\varphi\mid^{2})^{2} $$
$$ = \frac{1}{2e^{2}\mid\varphi_{c}\mid^{2}}(\partial_{\nu}\Lambda^{\nu\mu})^{2}+\frac{1}{4}e^{2}\{ -4\varphi_{c}^{2}(\partial_{\mu}\varphi_{2})^{2} 
     +\cdots \}
\qquad\qquad\qquad\qquad
$$
\begin{equation}
+ \frac{1}{2}\{ (\partial_{\mu}\varphi_{1})^{2}+(\partial_{\mu}\varphi_{2})^{2} \}+ \lambda \{ \varphi_{c}^{4} + 6\varphi_{c}^{2}(\varphi_{1}^{2}+\frac{2}{3}\varphi_{c}\varphi_{1}) + 2\varphi_{c}^{2}\varphi_{2}^{2} + \cdots \} ,
\end{equation}
where $\varphi_{1,2}$ is defined by
\begin{equation}
 \delta\varphi=\varphi_{1} + i\varphi_{2}, 
\end{equation}
and we used the relation, 
$\Lambda^{\mu\nu}=-\Lambda^{\nu\mu}$, in this expansion.
Then, the effective action has the form
\begin{equation}
 \Gamma_{4}(\varphi_{c})= - {\rm ln}\int {\cal D}\delta\varphi {\cal D}\delta\varphi^{\dagger} {\cal D}v {\cal D}\Lambda {\rm exp}\{ - \int d^{4}x  \tilde{\cal L}(\varphi_{c}+\delta\varphi) \}. 
\label{eqn:Effective Action}
\end{equation}
Substituting the Lagrangian density (\ref{eqn:Lagrangian}) into eq.(\ref{eqn:Effective Action}) and expanding $\tilde{\cal L}(\varphi_{c}+\delta\varphi)$ with respect to $\delta\varphi$, the effective action is given as
$$ \Gamma_{4}(\varphi_{c})= - {\rm ln}
\int {\cal D}\varphi_{1} {\cal D}\varphi_{2} {\cal D}v {\cal D}\Lambda {\rm exp}[ - \int d^{4}x \{ \frac{1}{4}v^{2}(\Lambda^{\mu\nu})^{2} + \frac{1}{2v^{2}}+\frac{1}{2}v^{2} $$
\begin{equation}
+ \frac{1}{2e^{2}\mid\varphi_{c}\mid^{2}}(\partial_{\nu}\Lambda^{\nu\mu})^{2}
+ \frac{1}{2}\bar{\varphi}_{1}(-\partial_{\mu}^{2}+12\lambda\varphi_{c}^{2})\bar{\varphi}_{1} + \varphi_{c}^{4} + 2\lambda\varphi_{c}^{2}\varphi_{2}^{2} \}].
\end{equation}
Here, we took only the terms corresponding to the one-loop contribution. 
After integrating $\varphi_{1}$ and $\varphi_{2}$, we obtain
$$ \Gamma_{4}(\varphi_{c})= - {\rm ln}
\int {\cal D}v {\cal D}\Lambda {\rm exp}\{ - \int d^{4}x [ \frac{1}{4}v^{2}(\Lambda^{\mu\nu})^{2} 
+ \frac{1}{2e^{2}\mid\varphi_{c}\mid^{2}}(\partial_{\nu}\Lambda^{\nu\mu})^{2}
+ \frac{1}{2v^{2}}+\frac{1}{2}v^{2} + \lambda\varphi_{c}^{4} $$
\begin{equation}
+ \frac{(12\lambda\varphi_{c}^{2})^{2}}{64\pi^{2}}( {\rm ln}\frac{12\lambda\varphi_{c}^{2}}{M^{2}}-1) ] \}, 
\end{equation}
where $M$ is some number with the dimensions of a mass.
Instead of $\Lambda^{\mu\nu}$, we define new fields by
\begin{equation}
 \bar{\Lambda}^{\mu\nu}=\frac{1}{\mid e\varphi_{c} \mid }\Lambda^{\mu\nu}, 
\end{equation}
then the action is rewritten as
$$ \Gamma_{4}(\varphi_{c})= - {\rm ln}
\int {\cal D}\varphi_{1} {\cal D}\varphi_{2} {\cal D}v {\cal D}\Lambda {\rm exp}\{ - \int d^{4}x [ \frac{1}{4}\gamma^{2}(\bar{\Lambda}^{\mu\nu})^{2} 
+ \frac{1}{2}(\partial_{\nu}\bar{\Lambda}^{\nu\mu})^{2}
+ \frac{1}{2v^{2}}+\frac{1}{2}v^{2} + \lambda\varphi_{c}^{4} $$
\begin{equation}
+ \frac{(12\lambda\varphi_{c}^{2})^{2}}{64\pi^{2}}[ {\rm ln}\frac{12\lambda\varphi_{c}^{2}}{M^{2}}-1] \}, 
\label{eqn:effective action2}
\end{equation}
where the parameter is defined as
\begin{equation}
 \gamma= \mid e\varphi_{c}v \mid . 
\end{equation}
The first and second terms of eq.(\ref{eqn:effective action2}) are expressed as
\begin{equation}
\frac{1}{4}\gamma^{2}(\bar{\Lambda}^{\mu\nu})^{2} + \frac{1}{2}(\partial_{\nu}\bar{\Lambda}^{\nu\mu})^{2}= \frac{1}{2}\bar{\Lambda}_{\mu\nu}D_{\mu\nu\rho\sigma}\bar{\Lambda}^{\rho\sigma},
\end{equation}
where $D_{\mu\nu\rho\sigma}$ is defined by the projection operators,
$\bar{P}^{(1)}_{\mu\nu\rho\sigma}$ and 
$\bar{P}^{(2)}_{\mu\nu\rho\sigma}$ as follows:
\begin{equation}
 D_{\mu\nu\rho\sigma}= -\frac{1}{2}\partial^{2}\bar{P}^{(1)}_{\mu\nu\rho\sigma} + \frac{1}{2}\gamma^{2}(\bar{P}^{(2)} + \bar{P}^{(1)})_{\mu\nu\rho\sigma}, 
\end{equation}
where
\begin{equation}
 \bar{P}^{(2)}_{\mu\nu\rho\sigma}=\frac{1}{2}(\theta_{\mu\rho}\theta_{\nu\sigma}-\theta_{\mu\sigma}\theta_{\nu\rho}), 
\end{equation}
\begin{equation}
 \bar{P}^{(1)}_{\mu\nu\rho\sigma}=\frac{1}{2}(
\theta_{\mu\rho}\omega_{\nu\sigma}-\theta_{\mu\sigma}\omega_{\nu\rho}
-\theta_{\nu\rho}\omega_{\mu\sigma}+\theta_{\nu\sigma}\omega_{\mu\rho}), 
\end{equation}
\begin{equation}
 \theta_{\mu\nu}= \delta_{\mu\nu}-\partial_{\mu}\partial_{\nu}/\partial^{2}, 
   \label{eq:th}
\end{equation}
\begin{equation}
\omega_{\mu\nu}= \partial_{\mu}\partial_{\nu}/\partial^{2}. 
   \label{eq:om}
\end{equation}
Integrating the field $\bar{\Lambda}^{\mu\nu}$, we obtain the effective action,
\begin{equation}
 \Gamma_{4}(\varphi_{c})= - {\rm ln}
\int {\cal D}v {\rm exp}\{ - \int d^{4}x U_{4}(\varphi_c,v) \}, 
\label{eqn:effective action}
\end{equation}
$$U_{4}= \frac{3\gamma^{4}}{64\pi^{2}}[{\rm ln}\frac{\gamma^{2}}{M^{2}}-1]
+ \frac{1}{2v^{2}}+\frac{1}{2}v^{2} + \lambda\varphi_{c}^{4} $$
\begin{equation}
+ \frac{(12\lambda\varphi_{c}^{2})^{2}}{64\pi^{2}}[ {\rm ln}\frac{12\lambda\varphi_{c}^{2}}{M^{2}}-a], 
\label{eqn:potential}
\end{equation}
where $M$ is some number with the mass-dimension, and $a$ is some number depending on a renormalization scheme.
Here, we assumed that $\gamma$ is slowly varying in the integration $\bar{\Lambda}^{\mu\nu}$.
If we assume that $\lambda$ is smaller than $e^{4}$, we can neglect the $\lambda$ and $\lambda^{2}$ terms in eq.(\ref{eqn:potential}). 
Then the potential can be written as
\begin{equation}
 U_{4}(u,v)= 
\frac{1}{2v^{2}}+\frac{1}{2}v^{2}
+ \frac{3u^{2}v^{4}}{64\pi^{2}}[{\rm ln}\frac{v^{2}u}{M^{2}}-1] , 
\end{equation}
where the parameter $u$ is defined by $u=e^{2}\varphi_{c}^{2}$.
Let us now forcus on the minimum of the potential $U_{4}(u,v)$ on the
$u-v$ plane. The minimum point $(u_{m},v_{m})$ is determined
by the following conditions
\begin{equation}
 \left. \frac{\partial U_{4}}{\partial\varphi_{c}} \right|_{u=u_{m},v=v_{m}}=0, \hspace{1cm} \left. \frac{\partial U_{4}}{\partial v} \right|_{u=u_{m},v=v_{m}}=0.
\label{eqn:minimum}
\end{equation}
From the above equations, the potential $U_{4}(u,v)$ has a minimum at $v_{m}=1.0$ and $u_{m}={\rm e}^{1/2}M^{2}$. 
Let us now see the potential $U_{4}$ as functions of $u$ and $v$. Fig.1 shows the potential surface of $U_{4}$ . 
In Fig.2, we can see that $U_{4}$ has a minimum near $v=1.0$. 
\begin{center}
      {\bf Fig.1, Fig.2}
\end{center}
It is difficult to integrate the auxiliary scalar field $v$ in eq.(\ref{eqn:effective action}), while we can integrate $v$ approximately by replacing it in the potential $U_{4}$ by its saddle point value $v_{s}$ which is determined by 
\begin{equation}
 \left. \frac{\partial U_{4}}{\partial v}\right|_{v=v_{s}}= -\frac{1}{v_{s}^{3}}+v_{s}+\frac{3uv_{s}^{3}}{16\pi^{2}}[{\rm ln}\frac{v_{s}^{2}u}{M^{2}}-\frac{1}{2}]=0. 
\label{eqn:saddle point}
\end{equation}
Fig.3 shows the curve of the saddle point. 
The value $v_{s}$ on the saddle point has a peak near $u=6$ and it decreases 
with increasing $u$. 
\begin{center}
     \bf{Fig.3}
\end{center}
Then, we can see that the saddle point (solid curve) intersects with the line of $v=1$ just at the minimum point $(u_{m},v_{m})$ of the potential $U_{4}$. 
Thus, the effective potential is derived as
\begin{equation}
 V_{4}(\varphi_{c})= 
\frac{1}{2v_{s}^{2}}+\frac{1}{2}v_{s}^{2}
+ \frac{3u^{2}v_{s}^{4}}{64\pi^{2}}[{\rm ln}\frac{v_{s}^{2}u}{M^{2}}-1] , 
\end{equation}
which has a minimum at ($u_{m}$,$v_{m}$), then the gauge symmetry is 
spontaneously broken. Here we notice that the effective potential is a funtion of $\varphi_{c}$ only since $v_{s}$ is represented by $\varphi_{c}$ from eq.(\ref{eqn:saddle point}).

  We show the effective potential as a function of $u$ in Fig.4. 
\begin{center}
   \bf{Fig.4}
\end{center}
If one takes $v=1.0$ in eq.(\ref{eqn:BIpart}) and the fourth order term of $\alpha'$ can be neglected in eq.(\ref{eqn:d=4}), then the BI action has simply the form, $F_{\mu\nu}F^{\mu\nu}$, which corresponds to the conventional action of a gauge theory. Then, the potential $U_{4}(v=1)$ is simply the Coleman-Weinberg potential \cite{cw}.
We now compare $V_{4}(\varphi_{c})$ with the potential $U_{4}(v=1)$. We cannot find difference between both potentials for $0\leq u \leq u_{m}$, 
while for $u > u_{m}$ the potential $V_{4}(\varphi_{c})$ gradually increases comparing with $U_{4}(v=1)$, and the difference is clear there. 
Thus, the higher order terms in the expansions of the BI action decrease the potential comparing with the conventional Coleman-Weinberg potential for 
$u > u_{m}$, that is, $V_{4}(\varphi_{c})<<U_{4}(v=1)$, while both 
potentials take the same minimum at the same point $u_{m}$.
We can see that this point, which the saddle point intersects with the
line of $v$=1, is just the minimum point of $V_4(\varphi_c)$.
Thus, we can see that the minimum of $V_{4}(\varphi_{c})$ coincides with one of $U_{4}(v=1)$, then the vacuum state would not change even if the BI action is used. 

\vspace{1cm}
\subsection{\bf $d=2$ and $3$}

  Let us next discuss the cases of $d=2$ and $3$. Following the same procedure as $d=4$ case, the effective action is given as
\begin{equation}
 \Gamma_{d}(\varphi_{c})= - {\rm ln}
\int {\cal D}v {\rm exp}\{ - \int d^{d}x U_{d} \}, 
\end{equation}
\begin{equation}
U_{2}= \frac{1}{2v^{2}}+\frac{1}{2}v^{2}-\frac{uv^{2}}{16\pi}[{\rm ln}\frac{uv^{2}}{M^{2}}-1],
\label{eqn:U2}
\end{equation}
\begin{equation}
U_{3}= \frac{1}{2v^{2}}+\frac{1}{2}v^{2}-\frac{u^{3/2}v^{3}}{3\sqrt{2}\pi}.
\label{eqn:U3}
\end{equation}
Here, we comment on the integration of $\Lambda^{\mu\nu}$ in the case of
$d=2$ and 3. For $d=2$, only one component of $\Lambda^{\mu\nu}$ is dynamical
variable.
Then we can write it as, $\Lambda^{\mu\nu}(x)=\epsilon^{\mu\nu}b(x)$, 
and its quadratic part 
in the action is given by
$$ -b[\partial^2-{1 \over 2}v^2(g\varphi_c)^2]b. $$
In the case of $d=3$, three components are dynamical variables 
and they are denoted 
by $B_{\mu}$, where 
$\Lambda^{\mu\nu}(x)=\epsilon^{\mu\nu\lambda}B_{\lambda}(x)$.
And their quadratic part is obtained as,
\[ -B^{\mu}\left\{[\partial^2-2v^2(g\varphi_c)^2]\theta_{\mu\nu}
          -{1 \over 2}v^2(g\varphi_c)^2\omega_{\mu\nu}\right\}
           B^{\nu}, \]
where $\theta_{\mu\nu}$ and $\omega_{\mu\nu}$ are given in (\ref{eq:th})
and (\ref{eq:om}) respectively. Except for these points, the procedure of
$d=4$ case is available.
   
Next, we show the results of the numerical analysis.
Fig.5 shows the potential surface of $U_{2}(u,v)$. It has a saddle. We show the curve of the saddle points in Fig.6. 
\begin{center}
   \bf{Fig.5, Fig.6}
\end{center}
This curve increases with increasing $u$ 
for large $u(>3)$ contrary to the previous case of $d=4$ shown in Fig.3. 
This can be easily understood from the sign of the third term of RHS in eq.(\ref{eqn:U2}). The corresponding part of $V_4$ has the opposite sign. In the case of $d=3$, the qualitative feature is similar to the case of $d=2$ as seen from Fig.5. However, the logarithmic term does not appear since the dimension is odd.

  The effective potentials in $d=2$ and 3 on the saddle point are 
\begin{equation}
V_{2}(\varphi_{c})= \frac{1}{2v_{s}^{2}}+\frac{1}{2}v_{s}^{2}-\frac{uv_{s}^{2}}{16\pi}[{\rm ln}\frac{uv_{s}^{2}}{M^{2}}-1],
\end{equation}
\begin{equation}
V_{3}(\varphi_{c})= \frac{1}{2v_{s}^{2}}+\frac{1}{2}v_{s}^{2}-\frac{u^{3/2}v_{s}^{3}}{3\sqrt{2}\pi}.
\end{equation}
The effective potential for $d=2$ is shown in Fig.7. 
\begin{center}
   \bf{Fig.7}
\end{center}
This effective potential has no minimum but a maximum. Therefore, the symmetry is not spontaneously broken in the cases of $d=2$ and 3. For $d=2$, no continuous symmetry breaking is consistent with the Coleman's theorem \cite{co}.
\vspace{1cm}

\section{Summary}

We have examined the effective potential of a scalar field
coupled with the gauge field whose action is supposed to be a high-energy 
form, i.e. the BI action. 
The BI action includes a series of $\alpha'$
expansion, so it is highly nonlinear with respect to the gauge fields.
However the minimum of the effective potential for d=4
coincides with the one obtained at low-energy where the form of the 
action of the gauge field is given by $F_{\mu\nu}F^{\mu\nu}$. 
Further, the maxima of the effective potential for $d=2,3$ are also the same
with that of low-energy one.
This result is correct
within the one-loop approximation with respect to the gauge coupling
constant, and the characteristic properties
of adopting the BI action will appear at higher orders of the gauge
coupling constant. While all the gauge coupling constants would be
asymptotic free, then they would be small at high energy. Then we
could say that the low-energy properties of the quantum effect
of gauge fields would not be changed even at high energy.
 
The above conclusion is useful when we consider an inflation model
at very early stage of the universe since we can use the Coleman-
Weinberg effective potential even there. The next problem is the 
consideration of the higher-order effect with respect to the 
string coupling constant $g_0$. Recently, many attempts in this
direction are tried from the viewpoint of the duality. This
is the beyond of our present work, and we will give it somewhere
else.

{}\baselineskip=10pt\parskip=0mm
\newpage
 
\newpage
{\bf Figure caption}
\begin{description}
\item{Fig.1} A three-dimensional plot of potential $U_{4}$.
\item{Fig.2} Counter map of potential $U_{4}$.
\item{Fig.3} Curve (solid line) of saddle point in $(u,v)$ plane for $d=4$.
\item{Fig.4} Effective potential $V_{4}(\varphi_{c})$ (solid line) on the saddle point and $U_{4}(v=1)$ (dotted line).
\item{Fig.5} Three-dimensional plot of potentials (a)$U_{2}$ and (b)$U_{3}$, and the counter map of potentials (c)$U_{2}$ and (d)$U_{3}$.
\item{Fig.6} Curve of saddle point in $(u,v)$ plane for $d=2$.
\item{Fig.7} Effective potential $V_{2}(\varphi_{c})$ on the saddle point.
\end{description}
\end{document}